\documentclass[conference]{IEEEtran}
\usepackage{graphicx}
\usepackage{booktabs}
\usepackage{makecell}
\usepackage{cite}
\usepackage{multirow}
\usepackage{multicol}
\usepackage{amsmath, amssymb, amsfonts}
\usepackage{textcomp}
\usepackage{xcolor}
\usepackage{url} 

\usepackage[caption=false]{subfig}  

\usepackage{algorithm}
\usepackage{algpseudocode}

\usepackage[colorlinks = true,
            linkcolor = blue,
            urlcolor  = blue,
            citecolor = blue,
            anchorcolor = blue]{hyperref}

\usepackage{etoolbox}

\usepackage{xcolor}

\author{\IEEEauthorblockN{Sandipan Dhar*, Mayank Gupta*, 	Preeti Rao}
    \IEEEauthorblockA{\textit{Department of Electrical Engineering, Indian Institute of Technology, Bombay.}}
    \IEEEauthorblockA{Email: \url{sandipandhartsk03@gmail.com}, \url{guptamayank8899@gmail.com}, \url{prao@ee.iitb.ac.in}}}

\begin{document}

\title{LAPS-Diff: A Diffusion-Based Framework for Singing Voice
Synthesis With Language Aware Prosody-Style Guided
Learning}


\maketitle

\begin{abstract}
The field of Singing Voice Synthesis (SVS) has seen significant advancements in recent years due to the rapid progress of diffusion-based approaches. However, capturing vocal style, genre-specific pitch inflections, and language-dependent characteristics remains challenging, particularly in low-resource scenarios. To address this, we propose LAPS-Diff, a diffusion model integrated with language-aware embeddings and a vocal-style guided learning mechanism, specifically designed for Bollywood Hindi singing style. We curate a Hindi SVS dataset and leverage pre-trained language models to extract word and phone-level embeddings for an enriched lyrics representation. Additionally, we incorporated a style encoder and a pitch extraction model to compute style and pitch losses, capturing features essential to the naturalness and expressiveness of the synthesized singing, particularly in terms of vocal style and pitch variations. Furthermore, we utilize MERT and IndicWav2Vec models to extract musical and contextual embeddings, serving as conditional priors to refine the acoustic feature generation process further. Based on objective and subjective evaluations, we demonstrate that LAPS-Diff significantly improves the quality of the generated samples compared to the considered state-of-the-art (SOTA) model for our constrained dataset that is typical of the low resource scenario.\footnote{* Sandipan Dhar and Mayank Gupta are co-first authors}
\end{abstract}


\section{Introduction}\label{sec:introduction}
Singing Voice Synthesis (SVS) is the process of generating natural-sounding singing voices using statistical or Artificial Intelligence (AI)-based algorithms, guided by musical scores \cite{Ref1-Diffsinger}. Unlike Text-to-Speech (TTS) synthesis systems that convert text into audible speech \cite{Ref2-Learn2Sing}, the primary objective of SVS models is to produce a singing voice that is accurately synchronized with the musical composition as given by a music score. Like a TTS model, an SVS model also follows a three-stage pipeline: converting the input music score into bottleneck features, generating acoustic features from bottleneck features, and reconstructing the audible singing voice \cite{Ref3-TCSinger}. In recent years, the growing use of SVS applications in the media and entertainment industry has highlighted the significant potential of this evolving field \cite{Ref2-Learn2Sing, Ref1-Singing-Tacotron-acm-2}.
\par
In the recent past, improved variants of diffusion models have exhibited significant advancements in SVS \cite{Ref4-MakeSinger, Ref5-HiddenSinger, Ref6-RDSinger, Ref1-Singing-ExpressiveSinger-acm-3}. In particular, the DiffSinger model \cite{Ref1-Diffsinger} is widely recognized as a pioneering approach in diffusion-based SVS research and has demonstrated promising results. It includes an auxiliary decoder that generates the target mel-spectrogram from music score embeddings. Specifically, the shallow diffusion mechanism introduced in DiffSinger accelerates the inference by providing the denoiser a boundary predicted intermediate mel-spectrogram rather than the noise input. Learn2Sing 2.0 \cite{Ref2-Learn2Sing} is also a well-known diffusion-based approach in which the authors adapted the GradTTS framework \cite{Ref19-Grad-TTS} to transform speech data into singing audio. In Learn2Sing 2.0, the authors effectively demonstrated how TTS models can be adapted for SVS with significant adjustments. Specifically, to improve the denoising process in the reverse diffusion for accurate estimation of the denoised mel-spectrogram having precise detail of the speaker's vocal characteristics, Learn2Sing 2.0~\cite{Ref2-Learn2Sing} introduced style embedding as an additional conditioning feature within the Grad-TTS framework. As a newer variant of diffusion-based approach, the authors of MakeSinger \cite{Ref4-MakeSinger} introduced an efficient semi-supervised training mechanism, enabling the model to learn from both labeled and unlabeled singing data. They conditioned the denoiser on text, pitch, and speaker embeddings, ensuring an improved reverse denoising process. While diffusion-based approaches have shown impressive results, models like VISinger2 \cite{Ref20-VISinger2} have also exhibited notable performance in SVS by utilizing a conditional variational autoencoder (CVAE)-based framework. 
\par
Although the discussed models have shown significant advancements, most of these existing SVS models, including DiffSinger, have been trained and evaluated on singing audio datasets \cite{Ref22-Opencpop, Ref21-Korean-Dataset, Ref24-JVS-MuSiC-Japanese, Ref23-Bisinger-English} that typically contain more than 5 hours of audio data (single voice) in languages such as Mandarin, Korean, Japanese, and English. Moreover, each of these language-specific singing datasets belongs to a distinct musical genre with its own peculiar characteristics for melody (pitch variation), rhythm, and pronunciation. These aspects are also strongly influenced by the individual singer's unique vocal characteristics. However, due to the scarcity of labeled singing data in most languages and genres, it is highly challenging for an SVS model to effectively capture linguistic content, style, and pitch related information from low-resource singing data, highlighting a significant research gap.
\par
To address this challenge of capturing content information, vocal style, and pitch variations from low-resource SVS data, 
we introduce LAPS-Diff, a diffusion model that incorporates language-aware embeddings and prosody-style guided learning mechanism. The proposed model is built on the DiffSinger framework \cite{Ref1-Diffsinger}. As part of this work, we curate a dataset of about one hour duration of Hindi Bollywood-style songs by a single male singer and process the audio to obtain the music scores. To effectively capture the lyrical content information, we leverage additionally two pre-trained language models, IndicBERT \cite{Ref11-IndicBERT} and XPhoneBERT \cite{Ref12-XPhoneBERT}, for Hindi word and phone-level embeddings. We combine these embeddings with music score embedding \cite{Ref1-Diffsinger} to form an enriched content representation. Further, we integrate a style encoder \cite{Ref14-StyleTTS2} and pre-trained JDCNet pitch extraction model \cite{Ref13-StyleTTS,Ref15-JDCNet} to obtain style and pitch information (melody) to compute the corresponding losses for the training of the auxilary decoder of our proposed LAPS-Diff model. Given the critical importance of preserving expressive pitch inflections in the songs in the genre of interest, we go beyond L1 loss of F0 (or log F0), and investigate the role of linear correlation in the loss function. The resulting style and pitch losses are found to improve the model’s ability to effectively capture vocal style and pitch-related information, enabling a closer resemblance to the underlying singing dynamics. Moreover, we utilize pre-trained MERT \cite{Ref17-MERT} and IndicWav2Vec \cite{Ref18-Indicwav2vec} models for extracting musical feature embeddings and contextual embeddings, respectively, to use as conditional priors to the denoiser for improving the reverse diffusion process \cite{Ref25-Diffusion-Process} through explicit feature guidance. Finally,
we evaluate the performances of the proposed LAPS-Diff model and the considered state-of-the-art (SOTA) SVS model using our Bollywood Hindi dataset with objective and subjective measures and discuss the obtained improvements. 
\section{Dataset}\label{sec:dataset}
Given that music generation must satisfy strong musical and cultural norms, it is critical to test with a dataset of the music in the genre of interest. We select the vastly popular genre of Indian music called Bollywood where iconic singers have large repertoires and significant following. While considered a pop genre in terms of song structure, Bollywood is known to be influenced by Indian classical traditions in terms of drawing on melodic material from ragas and folk songs, with Hindi as its primary language\footnote{Dataset details, and samples of reference and generated audio, can be accessed at anonymised site: \url{https://shorturl.at/2uARh}}.
We choose songs by popular singer Arijit Singh over the period 2012-2020. The selected 37 songs are in a single male voice.

\begin{figure}[htbp]
    \centering
    \includegraphics[width=\linewidth]{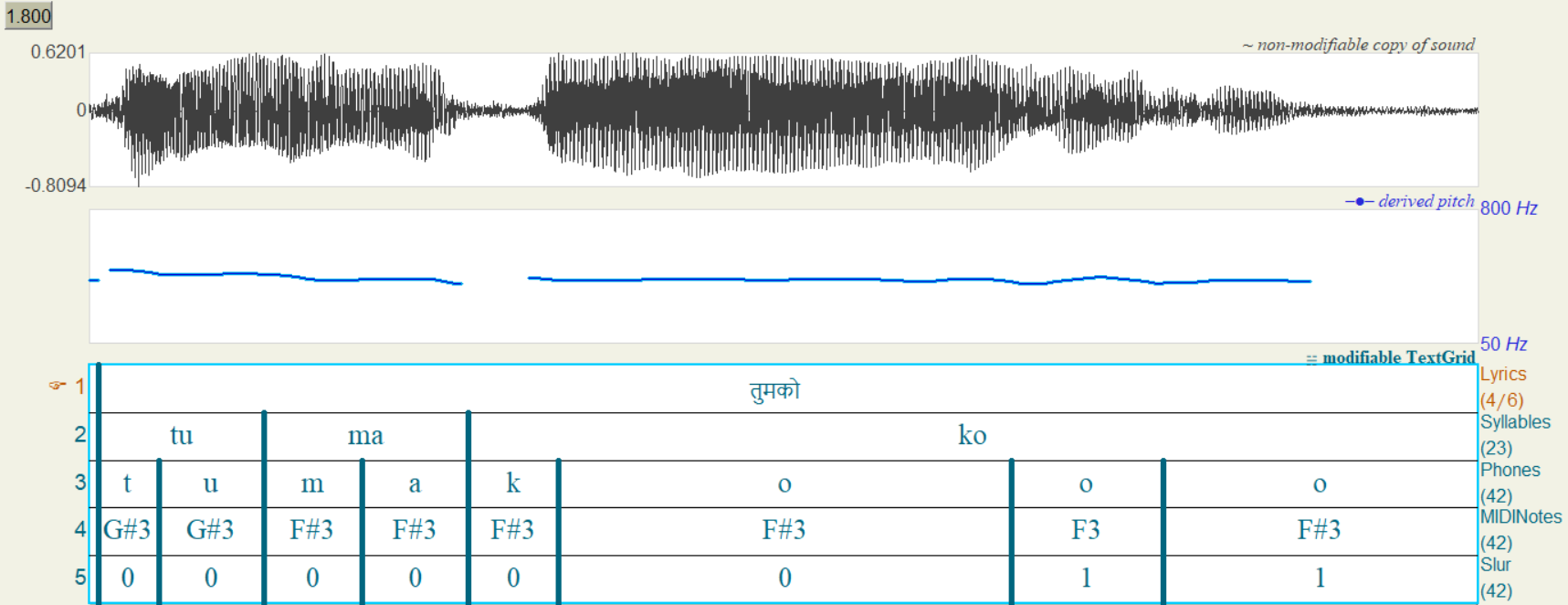}
    \vspace{-0.33cm}
    \caption{Waveform, pitch track and music score for a song segment from our dataset.}
    \label{Fig-2:txtgrid}
    \vspace{-0.33cm}
\end{figure}

The original song audio (as obtained from publicly accessible sites) is processed for vocals separation using a commercial tool \cite{Ref-GaudioLab}.
Automatically detected silences greater than 500 ms are used to segment the song into the lyrics lines/phrases that are then manually assigned to each audio segment. Next, we achieve the automatic alignment of the audio and text at the phone level using forced alignment with a hybrid automatic speech recognition system \cite{povey2011kaldi} that has been trained on 180 hours of adult Hindi read speech across 400 male and female speakers \cite{IITM_ASR_Challenge}. Sung lyrics exhibit notable differences from speech, chiefly on word pronunciations in terms of vowel duration as well as the nature of pitch variation over this duration. It is also common to find the addition of schwa on the last consonant of words. We account for these differences with a specially constructed lexicon by the systematic augmentation with alternate pronunciations. Syllable boundaries are obtained by merging the corresponding aligned phones. 
The vocal pitch is extracted at 10 ms intervals using an autocorrelation based method for fundamental frequency and voicing \cite{jadoul2018introducing}. Brief uv segments and pauses are linearly interpolated in pitch. Each syllable is assigned a MIDI note corresponding to the quantized mode of the F0 values across the syllable segment. Syllables longer than 200 ms are analysed for pitch fluctuations by computing the MIDI pitch separately for non-overlapping sub-segments of 200 ms and assigning a Slur flag to those segments that register a MIDI note change from the previous. This is the first attempt to the best of our knowledge to build an Indian music dataset for an SVS task. We adopt the same music score format as the Opencpop dataset \cite{Ref22-Opencpop} with lyrical content in text, corresponding phoneme sequence, associated musical notes, note duration, phoneme duration, and slur information, as depicted in Figure \ref{Fig-2:txtgrid}.

\section{Proposed Method}\label{sec:proposed-method}
The proposed model is built upon the DiffSinger framework \cite{Ref1-Diffsinger}. The same encoder model $E(.)$ is utilized to extract the music score embedding ${\boldsymbol{e}_{m}}$ from the music score $x$, as depicted in Fig. \ref{Fig-1:LAPS-Diff-model}. Similar to DiffSinger, the LAPS-Diff model incorporates two key components: the auxiliary decoder (aux decoder) and the denoiser. Training details of the auxiliary decoder and the denoiser, as well as the inference stage, are discussed in the following subsections.

\subsection{Feature Fusion}

\begin{figure*}[t]
    \centering
    \includegraphics[width=0.85\linewidth]{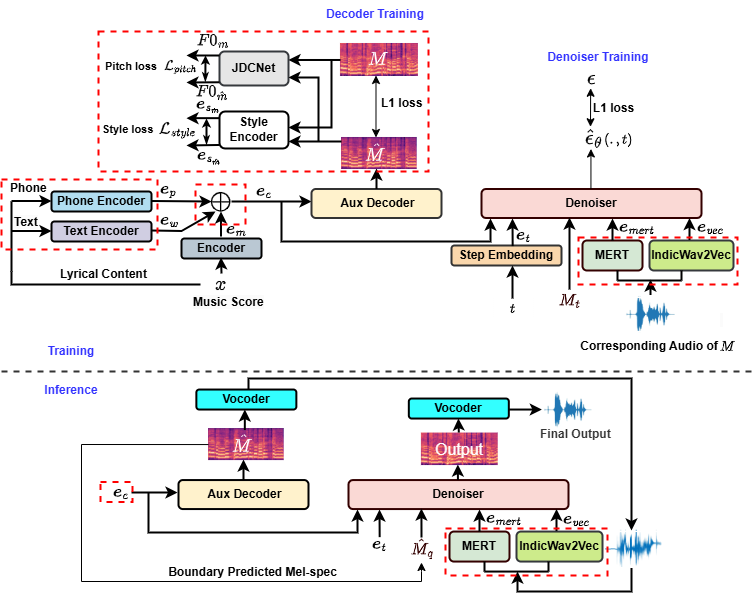}
    \vspace{-0.3cm}
    \caption{Schematic overview of the proposed LAPS-Diff model, including training and inference stages. The components enclosed within the red-dashed boxes represent the specific enhancements of this work over the DiffSinger framework. 
    }
    \label{Fig-1:LAPS-Diff-model}
    \vspace{-0.13cm}
\end{figure*}
To efficiently extract Hindi lyrical content related information from the textual content of the music score by obtaining an enriched lyrical representation, we use two pre-trained language models: IndicBERT (as a text encoder $T(.)$) and XPhoneBERT (as a phone encoder $H(.)$) both of which are compatible with the Hindi language. Thereafter, we combine IndicBERT's word level embedding ${\boldsymbol{e}_{w}}$ and XPhoneBERT's phone level embedding ${\boldsymbol{e}_{p}}$ with ${\boldsymbol{e}_{m}}$ to derive the final fused embedding ${\boldsymbol{e}_{c}}$ (i.e., fusion of three feature embeddings via summation operation \cite{Ref33-Summation-of-embeddings,Ref34-Summation-of-embeddings-1}), as described mathematically in Eq.(\ref{Eq:1})
(shown in Fig. \ref{Fig-1:LAPS-Diff-model}),
\vspace{-0.5mm} %
\begin{equation}
\label{Eq:1}
\begin{aligned}
    \mathbf{e}_c &= T(\text{text}) + H(\text{phone}) + E(x), \\
                 &= \mathbf{e}_w + \mathbf{e}_p + \mathbf{e}_m,
\end{aligned}
\end{equation}
The objective of constructing the fused embedding ${\boldsymbol{e}_{c}}$ is to provide a better representation of linguistic content (Hindi) in the embedding space, rather than relying solely on ${\boldsymbol{e}_{m}}$ as considered in DiffSinger.
\par

\par
\subsection{Decoder Training}
 The auxiliary decoder $AD(.)$ used in this work is a mel-spectrogram decoder capable of reconstructing mel-spectrograms $\hat M$ from the music score embedding \cite{Ref1-Diffsinger}. However, in this work, we provide the combined embedding ${\boldsymbol{e}_{c}}$ as the input to the auxiliary decoder to generate mel-spectrograms (i.e., $\hat{M} = AD(\boldsymbol{e}_c)$, where $\hat{M}$ is the generated version that is conditioned on the supplied embedding. In addition to content information, singer's vocal style and pitch variations carry the recognizable characteristics associated with any genre of singing. Hence, to effectively train the auxiliary decoder in capturing relevant information, we incorporate style and pitch losses as optimization objectives in our work. 
 \par
 \subsubsection{Style Loss}
 To extract the vocal style related information in terms of style embedding (style vector) from mel-spectrogram, we used a residual connection based Convolutional Neural Network (CNN) model as the style encoder $S(.)$ (the architectural framework of $S(.)$ is similar to the acoustic style encoder of StyleTTS2 \cite{Ref14-StyleTTS2}). The extracted style embedding carries an abstract representation (i.e., latent representation) of the singer's vocal characteristics. To compute the style loss $\mathcal{L}_{{style}}$, we first extract the style embeddings $\boldsymbol{e}_{s_{m}}$ and $\boldsymbol{e}_{s_{\hat m}}$ from both $M$ and $\hat{M}$ respectively, as illustrated in Fig. \ref{Fig-1:LAPS-Diff-model}. We then obtain $\mathcal{L}_{{style}}$ using the Mean Squared Error (MSE), as defined in the following equation,
\vspace{-1mm} %
\begin{equation}
\label{Eq:2}
\begin{aligned}
    \mathcal{L}_{{style}} &= \frac{1}{N} \sum_{i=1}^{N} \left\| \boldsymbol{e}_{s_{{m}_{i}}} - \boldsymbol{e}_{s_{\hat{m}_{i}}} \right\|^2,
\end{aligned}
\end{equation}
where $N$ is the total length of the embedding. In Eq.(\ref{Eq:2}),  $\boldsymbol{e}_{s_{m}}=S(M)$ and $\boldsymbol{e}_{s_{\hat m}}=S(\hat M)$.
\subsubsection{Pitch Loss}
The pitch loss $\mathcal{L}_{{pitch}}$ is calculated using MSE between the pitch values ${{F0}}_{m}$ and ${{F0}}_{\hat m}$ extracted from both $M$ and $\hat{M}$ respectively, using the pre-trained JDCNet model \cite{Ref15-JDCNet} denoted as $J(.)$. The pre-trained JDCNet model used in this work is adopted from StyleTTS2 \cite{Ref14-StyleTTS2}). Due to StyleTTS2's impressive voice cloning abilities in terms of naturalness and expressiveness, we kept the framework of $S(.)$ and $J(.)$ the same as in StyleTTS2. Moreover, JDCNet \cite{Ref15-JDCNet} is extensively trained on singing voice data, making it highly effective in capturing fine pitch details from mel-spectrograms \cite{Ref13-StyleTTS}. In a novel approach in this work, we introduce the concordance correlation coefficient (CCC) \cite{Ref34-correlation-coefficient} to exploit the linear correlation between matched pitch contours ${{F0}}_{m}$ and ${{F0}}_{\hat m}$. That is, the CCC (given in the Eq.(\ref{Eq:3}) below) balances the requirements of pitch height and correlation, making it particularly suitable for evaluating the perceptual similarity of two pitch contours. The CCC value ranges from $-1$ to $1$. To influence the pitch loss, $\mathcal{L}_{{pitch}}$, we incorporate $(1 - \text{CCC})$ as a multiplicative factor such that a highly correlated pair of ${{F0}}_{m}$ and ${F0}_{\hat{m}}$ results in a lower penalty, whereas a lower correlation increases the penalty.
The pitch loss $\mathcal{L}_{{pitch}}$ is mathematically defined in Eq.(\ref{Eq:3}),
\vspace{-5.4mm} %

\begin{gather}
\mathcal{L}_\text{pitch} = 
(1 - \text{CCC}) \times 
\left( \frac{1}{K} \sum_{i=1}^{K} \left\| F0_{m_i} - F0_{\hat{m}_i} \right\|^2 \right) \label{Eq:3} \\
\text{where} \quad 
\text{CCC} = 
\frac{
2\,\text{$\rho$}_{\scriptscriptstyle F0_m\, F0_{\hat{m}}}  
\sigma_{\scriptscriptstyle F0_m} 
\sigma_{\scriptscriptstyle F0_{\hat{m}}}
}{
\sigma_{\scriptscriptstyle F0_m}^2 + 
\sigma_{\scriptscriptstyle F0_{\hat{m}}}^2 + 
\left( \mu_{\scriptscriptstyle F0_m} - \mu_{\scriptscriptstyle F0_{\hat{m}}} \right)^2
} \raisebox{0.4ex}{\scalebox{1.1}{,}}
\nonumber
\end{gather}
and $\rho$ denotes the Pearson correlation coefficient between the predicted and ground-truth pitch contours, 
$\sigma$ represents the standard deviation, and $\mu$ denotes the mean value of the respective sequences. In Eq.(\ref{Eq:3}), $K$ represents the length of the $F0$ sequence, ${{F0}}_{m}=J(M)$ and ${{F0}}_{\hat m}=J(\hat M)$. To prevent the loss from becoming zero for highly correlated pairs, we introduce a minimum multiplicative factor of $0.01$.
\par
Besides $\mathcal{L}_{{style}}$ and $\mathcal{L}_{{pitch}}$, we incorporate $L_1$ loss (MAE loss) to measure the difference between $M$ and $\hat{M}$, following a similar approach to DiffSinger. Apart from these three losses, all other losses remain the same as in DiffSinger's Github repository \cite{DiffSinger-code}.
\subsection{Denoiser Training}
In the training phase of the denoiser $D(.)$ in DiffSinger, the model takes a noisy mel-spectrogram $M_{t}$ from the $t$-{th} ($t \in {0,T}$) diffusion step, along with the step embedding $\boldsymbol{e}_t$ for time step $t$ and the music score embedding $\boldsymbol{e}_m$, to estimate the noise $\hat{\epsilon}_{\theta}(\cdot)$ \cite{Ref1-Diffsinger} for obtaining $(t-1)$-{th} denoised melspectrogram $M_{t-1}$ as given in Eq.(\ref{Eq:4}),
\begin{equation}
\label{Eq:4}
M_{t-1} = \frac{1}{\sqrt{\alpha_t}} 
\left( M_t - \frac{1 - \alpha_t}{\sqrt{1 - \bar{\alpha}_t}} \hat{\epsilon}_{\theta}(M_t, \boldsymbol{e}_m, \boldsymbol{e}_t) \right) + \sigma_t z.
\end{equation}
In Eq.(\ref{Eq:4}), $\alpha_t = 1 - \beta_t$ ($\beta_t$ is variance schedule at time step $t$), $\bar{\alpha}_t = \prod_{s=1}^{t} \alpha_s$, $z \sim \mathcal{N}(0, I)$ and $\sigma_t$ is noise scale at time step $t$ \cite{Ref25-Diffusion-Process}.
\par
To improve the denoising process in reverse diffusion, particularly for accurate estimation of the denoised mel-spectrogram having precise detail of the musical and linguistic information, we introduce two additional feature embeddings as conditional priors to provide more explicit guidance during denoising. Between the two, one captures musical characteristics through MERT embedding \cite{Ref17-MERT} (focusing on tonal and pitch-related attributes), and the other captures linguistic information in terms of IndicWav2Vec \cite{Ref18-Indicwav2vec} embedding to obtain a richer representation in the denoised mel-spectrogram. MERT and IndicWav2Vec are two pre-trained models (trained using self-supervised learning approach) that efficiently extract respective embeddings from singing audio data (shown in Fig. \ref{Fig-1:LAPS-Diff-model}), as they are trained on singing (music) data and multilingual speech data (including Hindi), respectively. Therefore, Eq.(\ref{Eq:4}) can be reformulated as follows:
\begin{align}
\label{Eq:5}
M_{t-1} &= \frac{1}{\sqrt{\alpha_t}} \bigg( 
M_t - \frac{1 - \alpha_t}{\sqrt{1 - \bar{\alpha}_t}} 
\hat{\epsilon}_{\theta}(M_t, \boldsymbol{e}_c, \boldsymbol{e}_t, 
\boldsymbol{e}_{{mert}}, \boldsymbol{e}_{{vec}}) \bigg) \nonumber \\
&\quad + \sigma_t z.
\end{align}
In Eq.(\ref{Eq:5}), $\boldsymbol{e}_{{mert}}$ and $\boldsymbol{e}_{{vec}}$ represents MERT and IndicWav2Vec embeddings respectively. The denoiser training loss used in our work is identical to that of DiffSinger~\cite{DiffSinger-code} (i.e., $L_1$ loss). The model architecture of $D(.)$ considered in LAPS-Diff is identical to the denoiser used in DiffSinger \cite{Ref1-Diffsinger}.

\subsection{Inference Stage}
In the inference stage, the optimal auxiliary decoder $AD^{*}(\cdot)$ generates the mel-spectrogram $\hat{M}$, which contains nuanced representations of the lyrical content, vocal style, and prosody. The pre-trained HiFi-GAN vocoder \cite{Ref33:HiFi-GAN} is then used to reconstruct the audio waveform from $\hat{M}$, from which the embeddings $\boldsymbol{e}_{{mert}}$ and $\boldsymbol{e}_{{vec}}$ are extracted. These embeddings along with $\boldsymbol{e}_{{c}}$, $\boldsymbol{e}_{{t}}$, and the intermediate boundary-predicted \cite{Ref1-Diffsinger} representation $\hat{M}_{q}$ are used to condition the optimal denoiser $D^{*}(\cdot)$, as illustrated in Fig.~\ref{Fig-1:LAPS-Diff-model} to generate the final output. Here, $\hat{M}_{q}$ is derived from $\hat{M}$ through the shallow diffusion mechanism of DiffSinger~\cite{Ref1-Diffsinger}, where $q$ denotes the shallow diffusion step. Boundary prediction mechanism is applied only at the first iteration of the reverse diffusion, rest are same as Eq.(\ref{Eq:5}).
\par
In the DiffSinger model, the inference stage primarily relies on the boundary-predicted mel-spectrogram to generate the final output. In contrast, the proposed LAPS-Diff model explicitly incorporates MERT and IndicWav2Vec embeddings that are derived from the reconstructed waveform output of the vocoder. 
The complete workflow of the proposed LAPS-Diff model is summarized in Algorithm~\ref{Algo:1}.
\begin{algorithm}[t]
\caption{Workflow of the Proposed LAPS-Diff Model}
\label{Algo:1}
\begin{algorithmic}[1]
\State \textbf{Input:} Text, Phoneme, and Music score; Ground-truth audio and mel-spectrogram $M$
\State \textbf{Output:} Trained models $AD^*(\cdot)$ and $D^*(\cdot)$

\vspace{1mm}
\State \textbf{Embedding Extraction}
\State Extract embeddings: $\boldsymbol{e}_w$, $\boldsymbol{e}_p$, $\boldsymbol{e}_m$

\vspace{1mm}
\State \textbf{Training Auxiliary Decoder}
\State Generate mel-spectrogram: $\hat{M} = AD(\boldsymbol{e}_c)$
\State Extract $\boldsymbol{e}_{s_m}$, $\boldsymbol{e}_{s_{\hat{m}}}$
\State Extract $F0_m$, $F0_{\hat{m}}$
\State Compute style loss: $\mathcal{L}_{\text{style}}$
\State Compute pitch loss: $\mathcal{L}_{\text{pitch}}$
\State Compute reconstruction loss: $\mathcal{L}_{\text{L1}} = \text{MAE}(M, \hat{M})$
\State Compute total auxiliary loss: $\mathcal{L}_{\text{aux}} = \mathcal{L}_{\text{style}} + \mathcal{L}_{\text{pitch}} + \mathcal{L}_{\text{L1}}$
\State Update $AD(\cdot)$ to minimize $\mathcal{L}_{\text{aux}}$

\vspace{1mm}
\State \textbf{Training Denoiser}
\For{each diffusion step $t$}
    \State Generate noisy mel-spectrogram: $M_t$
    \State Get timestep embedding: $\boldsymbol{e}_t$
    \State Extract conditional MERT features: $\boldsymbol{e}_{\text{mert}}$
    \State Extract conditional Wav2Vec features: $\boldsymbol{e}_{\text{vec}}$
    \State Estimate noise: $\hat{\epsilon}_\theta = D(M_t, \boldsymbol{e}_c, \boldsymbol{e}_t, \boldsymbol{e}_{\text{mert}}, \boldsymbol{e}_{\text{vec}})$
    \State Compute denoising loss: $\mathcal{L}_{\text{denoise}} = \text{MAE}(\epsilon, \hat{\epsilon}_\theta)$
    \State Update $D(\cdot)$ to minimize $\mathcal{L}_{\text{denoise}}$
\EndFor
\State Obtain optimal models: $AD^*(\cdot)$ and $D^*(\cdot)$

\vspace{1mm}
\State \textbf{Inference Stage}
\State Extract embeddings as in previous steps
\State Generate initial mel-spectrogram: $\hat{M} = AD^*(\boldsymbol{e}_c)$
\State Perform shallow diffusion:
\State \hskip1em $\hat{M}_q = \text{ShallowDiffusion}(\hat{M}, \epsilon), \quad \epsilon \sim \mathcal{N}(0, I)$
\State Extract conditional MERT features: $\boldsymbol{e}_{\text{mert}}$
\State Extract conditional Wav2Vec features: $\boldsymbol{e}_{\text{vec}}$
\State Initialize diffusion mel: $M_q = \hat{M}_q$

\For{$t = q$ to $1$}
    \If{$t = 1$}
        \State $z = 0$
    \Else
        \State Sample $z \sim \mathcal{N}(0, I)$
    \EndIf
    \State Compute denoised mel-spectrogram:
    \State \hskip1em $
\begin{aligned}
M_{t-1} =  {} & \frac{1}{\sqrt{\alpha_t}} \Bigg( M_t - \frac{1 - \alpha_t}{\sqrt{1 - \bar{\alpha}_t}}  \hat{\epsilon}_\theta(M_t, \boldsymbol{e}_c, \boldsymbol{e}_t, \\
& \boldsymbol{e}_{\text{mert}}, \boldsymbol{e}_{\text{vec}}) \Bigg) + \sigma_t z
\end{aligned}$
\EndFor
\end{algorithmic}
\end{algorithm}

\vspace{-0.23cm}
\section{Experiments}
\label{experiment}
We present our experimental setup with implementation details. 

\subsection{Data Preprocessing}
The overall duration of the singing voice segments of our Hindi Bollywood SVS dataset, introduced in Section \ref{sec:dataset}, is approximately 65 minutes. With 38 unique songs, segmented into 397 sung phrases each with duration ranging from 5 to 15 seconds. The dataset splits are based on distributing the songs across training, validation, and test sets with details shown in Table \ref{tab:dataset_stats}. Each audio file is sampled at $16$ kHz with $16$-bit quantization. The Hindi text (Devanagari script) is converted into its corresponding phonemes using the IIT-M Hindi phoneset \cite{Ref33:IIT-M-Phoneset}. The size of the phoneme vocabulary considered is $43$. We extract mel-spectrograms considering $80$ mel-bins using frame size of $512$ and hop size of $128$. Similar to DiffSinger \cite{Ref1-Diffsinger}, the mel-spectrograms are normalized to the range $[-1, 1]$.

\begin{table}[t]
\centering
\caption{Description of dataset splits}
\vspace{-3mm}
\label{tab:dataset_stats}
\small
\resizebox{0.7\linewidth}{!}{%
\begin{tabular}{lcccc}
\toprule
\textbf{Split} & \textbf{Songs} & \textbf{Segments} & \textbf{Duration (Min)} \\
\midrule
Train       & 31 & 344 & 57.14 \\
Validation  & 3  & 25  & 3.76 \\
Test        & 4  & 28  & 3.76 \\
\bottomrule
\end{tabular}%
}
\vspace{-3mm}
\end{table}
\subsubsection{Implementation Details} In our work, we use the IndicBERT model \cite{Ref11-IndicBERT}, pre-trained on $12$ Indian languages, including Hindi. In contrast, the pre-trained XPhoneBERT model \cite{Ref12-XPhoneBERT}, which follows the BERT-base architecture \cite{Ref33-Summation-of-embeddings}, is trained on nearly $100$ languages, also including Hindi. The embedding dimension for both IndicBERT and XPhoneBERT is $768$. Linear projection is applied to reduce the dimensionality to 256, ensuring compatibility with the $256$-dimensional music score embedding $\boldsymbol{e}_{m}$. We retain the encoder architecture $E(.)$ as used in DiffSinger \cite{DiffSinger-code}. 
\par
The style encoder $S(.)$ consists of four 2D residual blocks \cite{Ref13-StyleTTS}, and the style vector dimensionality is set to 48. We use the checkpoint of the pre-trained JDCNet (a ConvBi-LSTM model consisting of three ResNet blocks) \cite{Ref15-JDCNet} from StyleTTS2 \cite{Ref14-StyleTTS2} to extract pitch values (the pitch sequence is of variable length depending on the segment duration). Unlike the loss functions described in the original DiffSinger paper \cite{Ref1-Diffsinger}, the official GitHub implementation \cite{DiffSinger-code} incorporates three additional duration losses—word, phone, and slur—duration along with a pitch and energy loss, employing duration, pitch and energy predictors similar to FastSpeech 2 \cite{FastSpeech_2}. These additional losses are also used to train the auxiliary decoder of both DiffSinger and our proposed model in all the experiments. The model architecture of $AD(.)$ is the same as FastSpeech2 \cite{Ref32-FastSpeech2, Ref1-Diffsinger}). In LAPS-Diff, the denoiser architecture $D(.)$ follows a non-causal WaveNet framework \cite{WaveNet}, similar to that used in DiffSinger \cite{Ref1-Diffsinger}. 
\par
The MERT model considered in our work is trained on diverse datasets comprising musical instruments and singing voices \cite{Ref17-MERT}. We use the pre-trained model to obtain the musical embeddings of size 
1024. The IndicWav2Vec \cite{Ref18-Indicwav2vec} is trained on nine Indian languages including Hindi, is capable of extracting linguistic content from speech audio. We employ the pre-trained IndicWav2Vec to obtain content embeddings of size 1024. We consider batch size as 48 and optimize the model using AdamW optimizer \cite{adamw} with a learning rate of $1\times{1}0^{-3}$, for a total training iteration of $2\times{10}^{5}$. Validation is performed in every $2\times{10}^{3}$ iterations. We set the value of $T = 100$ and  ${\beta}$ value increases linearly from from  $1\times{10}^{-4}$ to $6\times{10}^{-2}$ over the diffusion steps. While the DiffSinger paper \cite{Ref1-Diffsinger} describes the computation of the shallow diffusion step $q$ using boundary prediction mechanism, the official GitHub implementation \cite{DiffSinger-code} sets $q = 60$, which value we have retained in our implementation.
\par
The LAPS-Diff experiments are implemented in python 3.8.20 using pytorch 2.4.1 and numpy 1.19.2. Librosa 0.8.0 is used for audio preprocessing. All experiments run on an NVIDIA A100 GPU with 80GB memory. The training process takes approximately 2 GPU days to complete for the proposed model with the major component for complexity coming from the extra pitch calculation at each training iteration from the JDCNet \cite{Ref15-JDCNet} model.

\vspace{-0.43cm}

\section{Results and Discussion}
\label{results}
The performance of the proposed LAPS-Diff model is compared with the SOTA DiffSinger model using both objective and subjective tests, on our Hindi Bollywood SVS dataset. Additionally, an ablation study is carried out to demonstrate the effectiveness of the individual components integrated into LAPS-Diff to adapt it for Hindi singing data, using the same evaluation methods. For objective evaluation, we have used Mel-Cepstral Distortion (MCD), logarithmic Fundamental Frequency Root Mean Square Error (logF0 RMSE), Mean Absolute Error (MAE), cosine similarity and voiced/unvoiced (V/UV) ratio \cite{Ref28-MCD,ECAPATDNN-Speaker-Similarity,Ref30-VUV-Ratio} as evaluation metrics. Additionally, we have utilized the metas's Audiobox Aesthetic Score (AES) \cite{Ref26-AAT} for perceptual evaluation. Whereas, for subjective evaluation, we have considered Mean Opinion Score (MOS) \cite{Ref27-MOS}.

Four ablation settings (as summarised in Table \ref{tab:objective_eval} and Table \ref{tab:mos_ci}) are considered in our experiments: (i) training DiffSinger from scratch with text and phone embeddings (Ablation 1), (ii) training DiffSinger from scratch with MERT and IndicWav2Vec features as conditional priors (Ablation 2), (iii) training DiffSinger from scratch with JDCNet-based CCC pitch loss (Ablation 3), and (iv) training DiffSinger from scratch with style loss (Ablation 4). All settings were trained using the same dataset, preprocessing steps, number of training iterations, and overall hyperparameters to ensure consistency.


\subsection{Objective Evaluation}
The objective evaluation metrics considered here serve distinct purposes: MCD captures timbral (spectral) similarity, logF0 RMSE evaluates pitch similarity, MAE measures overall (global) structural similarity between mel-spectrograms, cosine similarity (with ECAPA-TDNN embeddings \cite{ECAPA-TDNN}) assesses the similarity of speaker dependent characteristics, and the V/UV ratio reflects the model’s ability to distinguish voiced and unvoiced regions. In contrast, AES provides a perceptual score using Meta’s pre-trained audio toolbox. Collectively, these metrics provide a comprehensive assessment of the generated samples across key acoustic dimensions.
\par

\begin{table*}[t]
\centering
\caption{Objective and perceptual evaluation of the proposed LAPS-Diff model, baseline DiffSinger, and its ablation variants}
\label{tab:objective_eval}
\renewcommand{\arraystretch}{1.2}
\resizebox{0.95\linewidth}{!}{%
\begin{tabular}{lccccccc}
\toprule
\textbf{Model} & 
\textbf{Cosine Similarity}~($\uparrow$) & 
\textbf{MAE}~($\downarrow$) & 
\textbf{V/UV Accuracy}~($\uparrow$) & 
\textbf{Log-$F_0$ RMSE}~($\downarrow$) & 
\textbf{MCD}~($\downarrow$) & 
\textbf{Audiobox CE}~($\uparrow$) & 
\textbf{Audiobox PQ}~($\uparrow$) \\
\midrule
Reference & -- & -- & -- & -- & -- & \textbf{6.206} & \textbf{7.637} \\
LAPS-Diff (Proposed) & \textbf{0.987} & \textbf{0.165} & \textbf{0.907} & 0.141 & 7.897 & 4.770 & 6.552 \\
DiffSinger (Baseline) & 0.982 & 0.197 & 0.890 & 0.155 & 8.200 & 4.004 & 6.340 \\
Ablation 1 (Baseline + IndicBERT + XPhoneBERT) & 0.973 & 0.171 & 0.890 & 0.159 & 7.983 & 4.200 & 6.499 \\
Ablation 2 (Baseline + MERT + IndicWav2Vec) & 0.978 & 0.185 & 0.869 & 0.151 & 9.445 & 4.151 & 6.408 \\
Ablation 3 (Baseline + JDCNet Pitch Loss) & 0.978 & 0.171 & 0.898 & \textbf{0.118} & 7.928 & 3.460 & 6.355 \\
Ablation 4 (Baseline + Style Loss) & 0.986 & 0.169 & 0.880 & 0.145 & \textbf{7.883} & 3.869 & 6.511 \\
\bottomrule
\end{tabular}%
}
\end{table*}

\par
The objective evaluation results in Table \ref{tab:objective_eval} are the average computed across all the test data. We note the superior performance of the proposed LAPS-Diff model compared to the baseline DiffSinger. LAPS-Diff outperforms across most metrics, achieving the highest average cosine similarity, lowest MAE, highest V/UV accuracy, second lowest log-F0 RMSE and MCD. These results suggest that LAPS-Diff captures speaker characteristics more accurately and ensures closer alignment of the spectral features (including pitch) with the ground truth, thereby enhancing the overall quality of the synthesized singing voice. Moreover, the higher V/UV ratio achieved by the proposed model indicates its effectiveness in capturing the details of voiced and unvoiced regions, contributing to greater naturalness in the synthesized singing voice.
\par
In Ablation 1, we retain only the content encoder enhanced part (i.e., $\boldsymbol{e}_c$), replacing all other LAPS-Diff innovations with their vanilla counterparts (i.e., DiffSinger). This configuration showed moderate performance gains over vanilla DiffSinger, in terms of MAE and MCD ratio. However, due to the absence of stylistic and prosodic control, the generated outputs lacked expressive nuance, evident from a higher log-F0 RMSE and reduced cosine similarity. The results show that the absence of prosody and style modeling greatly hampers the melodic and stylistic quality of the synthesized singing voice. In Ablation 2 variant, only the MERT and IndicWav2Vec priors were added to denoiser, while other LAPS-Diff components were excluded. This setup led to notable improvements in MAE, MCD, and log-F0 RMSE compared to the vanilla DiffSinger, suggesting that these embedding priors effectively enhance spectral feature generation. However, the cosine similarity and the V/UV ratio decreased, highlighting the necessity of style and prosody embeddings for better speaker similarity and accurate modeling of voiced/unvoiced regions. In Ablation 3 the JDCNet-based pitch loss retained, excluding all other LAPS-Diff components. By introducing pre-trained prosody-level supervision, the model achieved notable improvements in log-F0 RMSE, MCD, MAE, and V/UV ratio reflecting more accurate and smoother pitch and spectral modeling. However, the cosine similarity slightly decreased compared to the vanilla DiffSinger. Finally, in Ablation 4 variant, only the style loss is retained by replacing all other components of LAPS-Diff. This enables explicit modeling of style enhanced expressive features as compared to vanilla DiffSinger. This setting demonstrated improvements in cosine similarity (as compared to vanilla DiffSinger) suggesting better stylistic alignment, but a decline in the V/UV ratio. However, it showed significant gains in terms log-F0 RMSE, MCD, and MAE. These results highlight that while style loss effectively introduces expressive variation, it alone is insufficient for managing the dynamic aspects of singing that influence V/UV regions.
\par
Meta’s AES evaluation metrics \cite{Ref26-AAT}, adopted in our work, are designed to assess the aesthetic quality of diverse audio types including speech, music and general sounds, using a pre-trained model that closely aligns with human perceptual judgments.
We select the measures relevant to synthesis from music score (i.e. content is pre-defined), viz.  Production Quality (PQ) capturing the technical clarity and fidelity of the audio, including its dynamics, frequency balance, and spatial attributes. 
Content Enjoyment (CE) evaluates the emotional and artistic expressiveness of the audio—highlighting aspects like creativity, mood, and subjective appeal. Since we have a single voice, the AES "production complexity" measure (PC) is irrelevant. 

\par
Table \ref{tab:objective_eval} clearly shows that the LAPS-Diff model outperforms all other models across both AES dimensions, achieving the highest scores (except for the reference scores, as expected) 
demonstrating its superior performance in generating singing voice samples that are highly expressive and perceptually rich. Among the ablation variants, Ablation 1  and Ablation 2 show notable improvements over the DiffSinger baseline, particularly in terms of CE and PQ, highlighting the importance of linguistic and musical embeddings in enhancing expressiveness and the overall quality of the generated samples. Additionally, Ablation 3 and Ablation 4 contribute primarily to technical fidelity, clarity, and harmonic richness of the generated singing voice (i.e., PQ) but fall slightly short in capturing artistic expressiveness (CE), suggesting that this attribute likely needs other dimensions such as timbre, loudness and voice quality. 
\par
\subsubsection{Visual Analysis}
\par
\begin{figure}[t]
    \centering
    \includegraphics[width=0.8\linewidth]{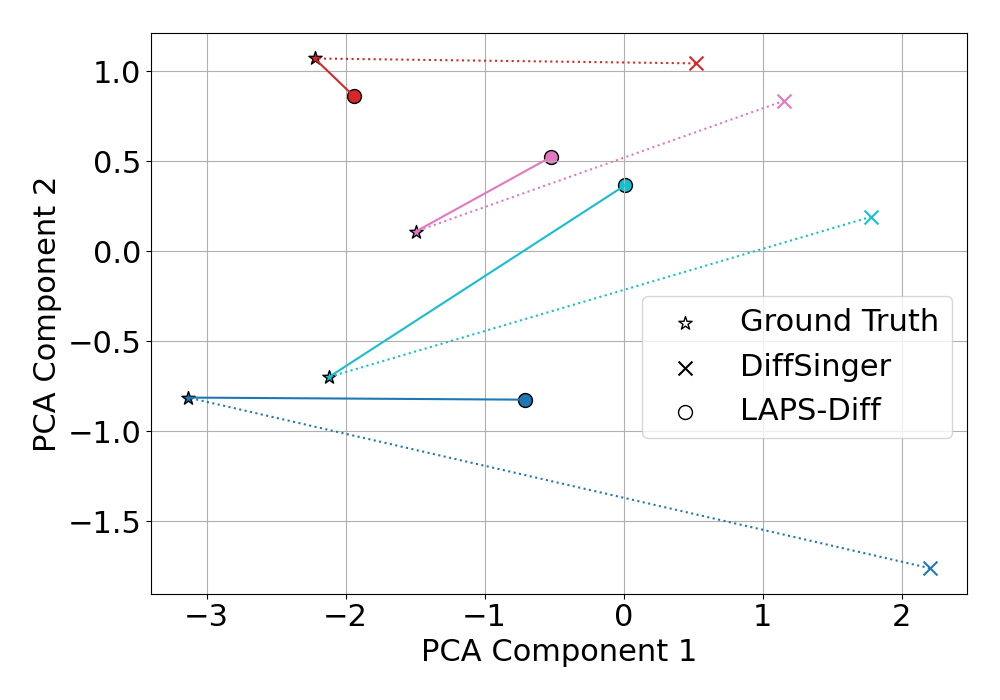}
    \vspace{-0.3cm}
    \caption{PCA visualization of content embeddings extracted using IndicWav2Vec from ground truth, DiffSinger, and LAPS-Diff outputs, illustrating content-level similarity. Here, each color represents a unique audio segment.}
    \label{Fig-2:PCA}
\end{figure}

To compare the lyrical content preservation capabilities of the proposed model and the baseline, we extracted Hindi content embeddings using IndicWav2Vec from LAPS-Diff, DiffSinger, and the ground truth audio. These high-dimensional embeddings are projected onto a 2D space using Principal Component Analysis (PCA) \cite{PCA} for visualization. As illustrated in Fig. \ref{Fig-2:PCA}, the LAPS-Diff embeddings are located closer to the ground truth than those from DiffSinger, indicating that LAPS-Diff captures content information more effectively.
\par
\begin{figure*}[t]
    \centering
    \includegraphics[width=0.33\textwidth]{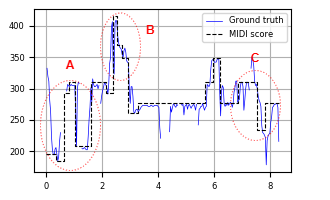}\hfill
    \includegraphics[width=0.33\textwidth]{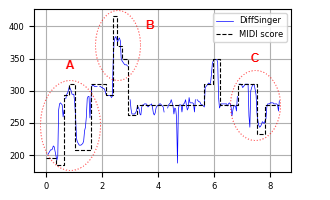}\hfill
    \includegraphics[width=0.33\textwidth]{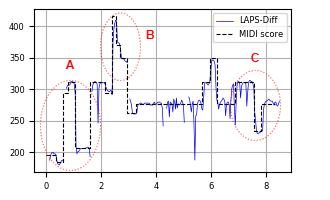} \\[1ex]
    \includegraphics[width=0.3\textwidth]{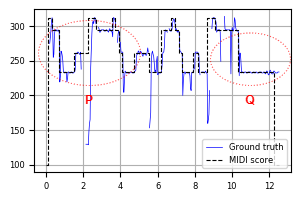}\hfill
    \hspace*{-0.5em}\includegraphics[width=0.3\textwidth]{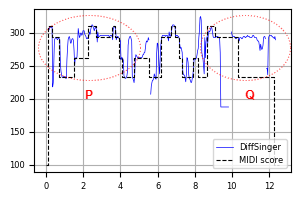}\hfill
    \includegraphics[width=0.3\textwidth]{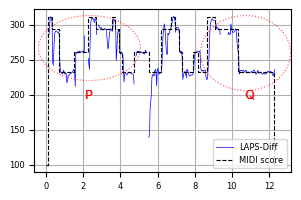}\hspace*{1.2em}

    \caption{Visualization of the F0 contour comparing ground truth with synthesized outputs from DiffSinger and LAPS-Diff, all with reference to the MIDI score. The vertical axis shows frequency (Hz), and the horizontal axis represents time (seconds). Top row contains a sample with faster singing rate, and bottom shows a sample with slower singing rate.}
    \label{fig:six_grid_f0}
\end{figure*}

Next, we present a visual comparison of pitch contours from selected audio segments for LAPS-Diff, DiffSinger and the reference, corresponding to faster and slower singing rates shown in Fig.~\ref{fig:six_grid_f0}. 
For the fast singing rate, the pitch contours in regions \( A \) and \( C \) of LAPS-Diff closely resemble those of the corresponding ground truth regions, whereas DiffSinger shows noticeable deviations. The region \( B\) contains a pitch transition to a much higher value, and Fig.~\ref{fig:six_grid_f0} shows LAPS-Diff appropriately synthesizes the high-pitched part, matching the ground truth. In contrast, DiffSinger's pitch pattern in region \( B\) deviates visibly from the music score MIDI. This indicates LAPS-Diff's ability in handling high-pitched regions and highlights its effectiveness in capturing fine-grained pitch details across diverse frequency ranges. A similar observation holds for region \( P \) (Fig.~\ref{fig:six_grid_f0}) in the slower singing rate scenario, where LAPS-Diff better matches the ground truth contours. Moreover, in region \( Q \), which corresponds to a held (and therefore long duration) vowel, DiffSinger fails to follow the within-vowel pitch variation, while LAPS-Diff shows high similarity with the ground truth. These comparisons indicate that LAPS-Diff captures pitch information more accurately and consistently than DiffSinger, which may be attributed to the additional correlation-dependent pitch loss.
\par
\begin{figure*}[t]
    \centering

    \begin{minipage}{0.33\textwidth}
        \centering
        \includegraphics[width=\textwidth]{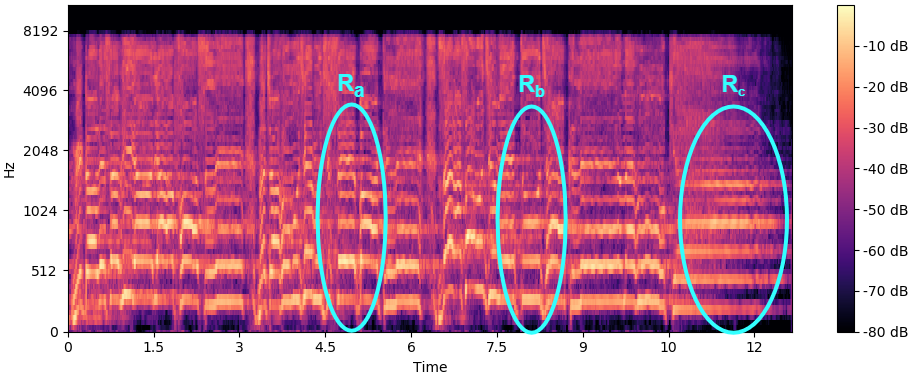} \\
        \includegraphics[width=\textwidth]{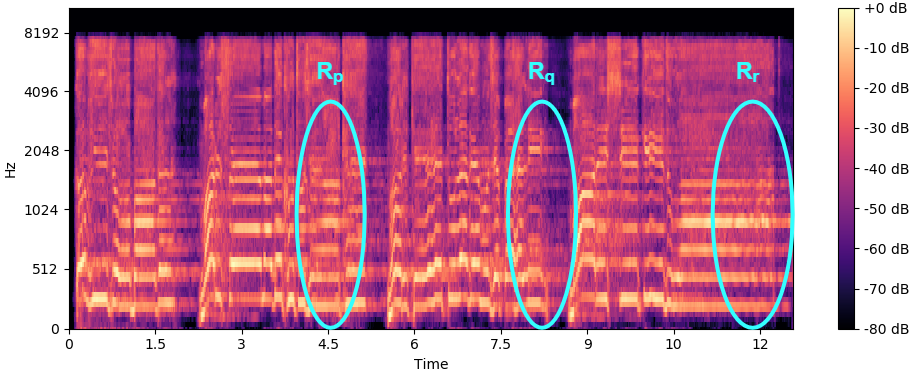} \\
        \textbf{Ground Truth}
    \end{minipage}%
    \hfill
    \begin{minipage}{0.33\textwidth}
        \centering
        \includegraphics[width=\textwidth]{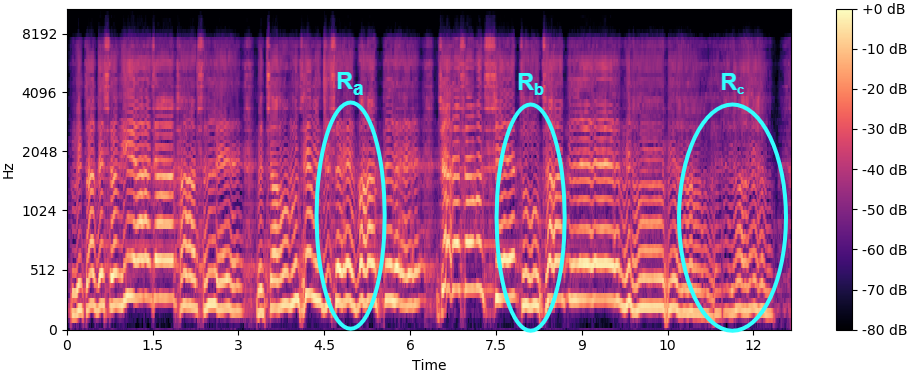} \\
        \includegraphics[width=\textwidth]{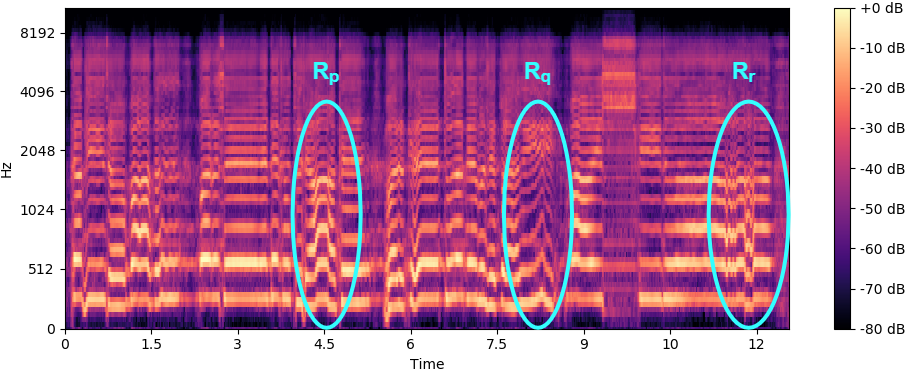} \\
        \textbf{DiffSinger}
    \end{minipage}%
    \hfill
    \begin{minipage}{0.33\textwidth}
        \centering
        \includegraphics[width=\textwidth]{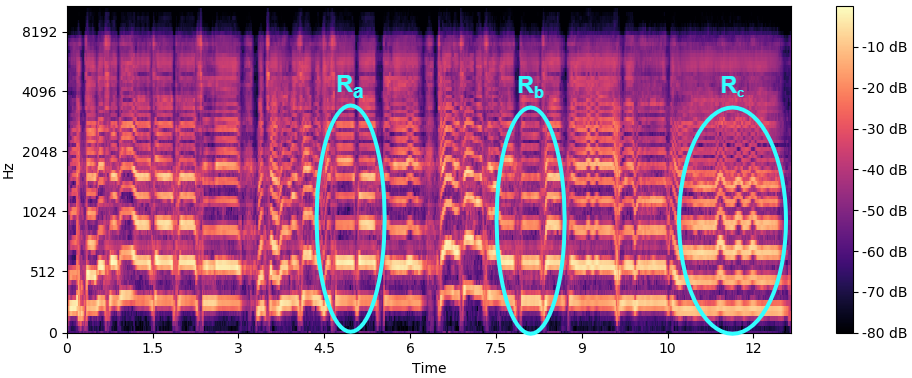} \\
        \includegraphics[width=\textwidth]{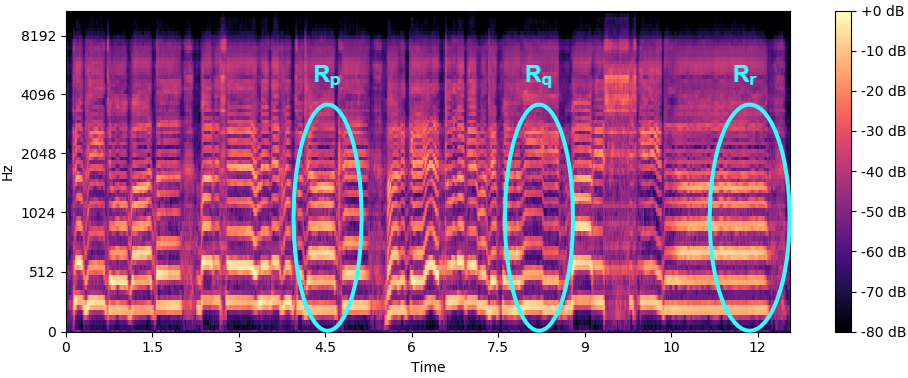} \\
        \textbf{LAPS-Diff}
    \end{minipage}

    \caption{Mel spectrograms comparing ground truth with synthesized outputs from DiffSinger and LAPS-Diff. Top row contains a sample with faster singing rate, whereas the bottom row features a sample with slower singing rate.}
    \label{fig:six_image_grid}
    \vspace{-0.43cm}
\end{figure*}

We also present a visual comparison of mel-spectrograms to evaluate the effectiveness of our proposed model. Two audio segments are considered: one with a faster singing rate and the other with a slower rate, shown in the top and bottom rows of Fig.~\ref{fig:six_image_grid}, respectively. The spectrograms are divided into three regions for detailed analysis. In the first row, corresponding to the faster singing rate, regions \( R_a \) and \( R_b \) exhibit significant harmonic distortion in DiffSinger, whereas LAPS-Diff preserves the harmonic structure more accurately. In region \( R_c \), an elongated phoneme appears noticeably distorted in DiffSinger, while LAPS-Diff renders it more clearly with minimal artifacts. In the second row of Fig.~\ref{fig:six_image_grid}, which corresponds to the slower singing rate, region \( R_p \) again shows harmonic distortion in the output of DiffSinger, while LAPS-Diff maintains clear harmonic patterns. In region \( R_q \), DiffSinger struggles to represent low-energy (unvoiced) segments, which are better captured by LAPS-Diff. Lastly, in region \( R_r \), stressed phonemes are distorted in DiffSinger, whereas LAPS-Diff preserves their clarity. These observations highlight LAPS-Diff's superior ability to model elongated phonemes and stylistic nuances, which are crucial for high-quality singing voice synthesis. Since the audio samples used in Fig.~\ref{fig:six_grid_f0} and Fig.~\ref{fig:six_image_grid} are not the same, their time durations differ.

\vspace{-0.33cm}
\subsection{Subjective Evaluation}
\subsubsection{MOS Testing}
We conducted MOS subjective evaluation test to assess the naturalness of the generated audio samples and the ground truth. A total of 15 listeners, aged between 20-35 years and without any known hearing impairments, participated. Each participant was asked to rate the samples on a 5-point scale \cite{Ref27-MOS}. Our test set consists of 4 songs: one each with fast and slow singing rates and two with average singing rates. The evaluation included $6$ segments (out of a total of 28) from our test set, ensuring that we take segments from each category of fast, slow, and average singing rate. We take segments from each of the following $7$ categories: ground truth, DiffSinger, LAPS-Diff (proposed model), and four ablation variants, resulting in a total of 42 samples. These were randomly permuted and arranged in $6$ test sets of $7$ each. The results of the MOS subjective evaluation test are shown in Table \ref{tab:mos_ci}.
\begin{table}[t]
\centering
\caption{MOS with 95\% confidence intervals}
\label{tab:mos_ci}
\resizebox{0.9\linewidth}{!}{%
\begin{tabular}{lc}
\toprule
\textbf{Model} & \textbf{MOS} ($\uparrow$) \\
\midrule
Reference             & \textbf{4.59 $\pm$ 0.26} \\
LAPS-Diff (Proposed)              & 3.36 $\pm$ 0.34 \\
DiffSinger (Baseline)    & 2.85 $\pm$ 0.44 \\
Ablation 1 (Baseline + IndicBERT + XPhoneBERT)               & 2.83 $\pm$ 0.58 \\
Ablation 2 (Baseline + MERT + IndicWav2Vec)              & 3.01 $\pm$ 0.18 \\
Ablation 3 (Baseline + JDCNet pitch loss)              & 3.05 $\pm$ 0.38 \\
Ablation 4 (Baseline + style loss)              & 2.95 $\pm$ 0.47 \\
\bottomrule
\end{tabular}%
}
\vspace{-3mm}
\end{table}
\par
The results of Table \ref{tab:mos_ci} clearly highlight the effectiveness of each component integrated into the proposed LAPS-Diff model. As observed from the MOS scores, the proposed LAPS-Diff model achieves a significantly higher MOS (3.36) compared to the baseline DiffSinger (2.85), confirming the benefit of our proposed enhancements. Ablation 1 results in a low MOS score, indicating that linguistic features alone are insufficient for adequately capturing the expressive qualities of the singing voice that are needed to enhance perceptual naturalness. Listeners reported that while this ablation has good content information, it has unwanted artifacts in pitch in long phones and instances with slurs, indicating the need for improvement in the pitch. Ablation 2, achieves a higher MOS (3.01), suggesting that musical and contextual embeddings help to guide the denoiser more effectively. Listeners reported smoother transitions across different phones and stated that the audio seemed more natural, as seen by the high PQ score in the perceptual evaluation in Table \ref{tab:objective_eval}. Similarly, Ablation 3 further improves the naturalness, revealing how accurate pitch modeling contributes to generating more melodically coherent singing voice. Listeners reported that the prosody and pitch quality in this ablation was significantly better than other ablations, but at the same time, intelligibility was reduced, indicating the necessity of combining all the contributions. Ablation 4 reaches 2.95, pointing out the role of vocal style in enhancing expressiveness. In this Ablation, listeners perceived the expected singing expressiveness and naturalness compared to the other cases. Overall, integrating linguistic embeddings, musical and contextual priors, along with style- and pitch-guided losses, results in a comprehensive performance improvement for the LAPS-Diff model, showcasing its capability to generate natural and expressive Hindi singing voices even under limited data conditions.
\vspace{-0.2cm}
\section{Conclusion}
\label{conclusion}
\par
In this work, we introduced LAPS-Diff, a novel singing voice synthesis model designed for low-resource Hindi Bollywood singing style. We present a new labeled dataset for Hindi Bollywood music of one-hour duration (audio provided via available public links). By incorporating enriched linguistic content embeddings, style and pitch supervision through loss computation, and utilizing conditional priors, LAPS-Diff achieves notable improvements in the expressiveness and fidelity of the synthesized singing voice over that of the DiffSinger baseline it is derived from. In particular, the implicit learning of expressive pitch contours is facilitated by the use of a new pitch loss that considers also the correlation of the pitch contours (i.e. melodic shape matching).  Our results, supported by both objective and subjective evaluations, confirm that incorporating pre-trained language models alongside singing voice-specific feature modules greatly improves the quality of SVS, as observed in the low-resource context targeted in this work. Future efforts will address the computational optimization of training and inference. We will study gains in generated quality that are achievable as the dataset size is expanded, including multilingual SVS, with an emphasis on enhancing generalization across a variety of vocal styles.

\bibliographystyle{IEEEtran}
\bibliography{Bibfile.bib}
\end{document}